\documentclass[twocolumn,tighten]{aastex63}

\usepackage{graphicx}
\usepackage{amsmath}
\usepackage{amssymb}
\usepackage{color}
\usepackage{mathtools}
\usepackage{ulem}
\usepackage[all]{hypcap} 

\usepackage{lineno}

\newcommand{\Msun}{\,{\rm M_\odot}}
\newcommand{\Mblack}{M_\bullet}

\begin{document}

\title{Constraining the Properties of Black Hole Seeds from the Farthest Quasars}

\author[0000-0002-7330-027X]{Giacomo Fragione}
\affil{Center for Interdisciplinary Exploration \& Research in Astrophysics (CIERA), Northwestern University, Evanston, IL 60208, USA}
\affil{Department of Physics \& Astronomy, Northwestern University, Evanston, IL 60208, USA}

\author[0000-0001-9879-7780]{Fabio Pacucci}
\affiliation{Center for Astrophysics $\vert$ Harvard \& Smithsonian, Cambridge, MA 02138, USA}
\affiliation{Black Hole Initiative, Harvard University, Cambridge, MA 02138, USA}

\begin{abstract}
Over 60 years after the discovery of the first quasar, more than $275$ such sources are identified in the epoch of reionization at $z>6$. JWST is now exploring higher redshifts ($z\gtrsim 8$) and lower mass ($ \lesssim 10^7 \Msun$) ranges. The discovery of progressively farther quasars is instrumental to constraining the properties of the first population of black holes (BHs), or BH seeds, formed at $z \sim 20-30$. For the first time, we use Bayesian analysis of the most comprehensive catalog of quasars at $z>6$ to constrain the distribution of BH seeds. We show that the mass distribution of BH seeds can be effectively described by combining a power law and a lognormal function tailored to the mass ranges associated with light and heavy seeds, assuming Eddington-limited growth and early seeding time. Our analysis reveals a power-law slope of $-0.70^{+0.46}_{-0.46}$ and a lognormal mean of $4.44^{+0.30}_{-0.30}$. The inferred values of the Eddington ratio, the duty cycle, and the mean radiative efficiency are $0.82^{+0.10}_{-0.10}$, $0.66^{+0.23}_{-0.23}$, and $0.06^{+0.02}_{-0.02}$, respectively. Models that solely incorporate a power law or a lognormal distribution within the specific mass range corresponding to light and heavy seeds are statistically strongly disfavored, unlike models not restricted to this specific range. Our results suggest that including both components is necessary to comprehensively account for the masses of high-redshift quasars, and that both light and heavy seeds formed in the early Universe and grew to form the population of quasars we observe.
\end{abstract}


\section{Introduction} \label{sec:intro}

Sixty years ago, the first quasar, cataloged as 3C 273, was identified at $z = 0.158$ \citep{Schmidt_1963}. At the time of discovery, this source was the farthest ever observed and led to a race to identify the mechanism responsible for such an efficient transformation of matter into energy (see, e.g., \citealt{Salpeter_1964}). Since then, quasar discoveries have exploded in number and consistently broken distance records. With the Sloan Digital Sky Survey (SDSS), quasars at $z\sim 5-6$ \citep{Fan_1999, Fan_2001}, and then well into the reionization epoch \citep{Fan_2006} were discovered.

Nowadays, we routinely detect quasars at $z \gtrsim 7$ \citep{Wang_2021}. According to a recent and comprehensive review, we have detected 275 quasars at $z > 6$ and 8 at $z > 7$ \citep{Fan_2022_review}. To date, the farthest supermassive black hole (SMBH) ever detected, GN-z11, was discovered in the JADES survey by the James Webb Space Telescope (JWST) at $z=10.6$ \citep{Maiolino_2023}, or only $440$ Myr after the Big Bang, assuming Planck cosmological parameters \citep{Planck_cosmo_2020}. Note that this SMBH, while being significantly farther than previous record-holders, is characterized by a mass of only $\log_{10} \Mblack = 6.2\pm 0.3 \Msun$, offering for the first time a glimpse of the lower end of the SMBH distribution at such redshifts.

Upcoming surveys with new observational facilities (e.g., Euclid, the Nancy Grace Roman Space Telescope) will extend our reach of searches and detect even farther quasars \citep{Tee_2023}. Current forecasts predict that the first data release of the near-infrared telescope Euclid will lead to $13-25$ new quasar discoveries in the redshift range $7 < z < 9$ \citep{Euclid_2019}. These forecasts are based on linear extrapolations at higher redshifts of the rate of decrease of the spatial density of quasars currently detected. Based on current estimates of the rate of decline, \cite{Fan_2019BAAS} and \cite{Wang_2019_density} predict that the farthest quasar with a mass $\Mblack > 10^9 \Msun$ should be detected in the redshift range $9 < z < 12$, depending on the average accretion rate characterizing these systems. 

Remarkably, in this redshift range, the JWST has already identified many galaxy candidates with photometric (e.g., \citealt{Castellano_2022, Labbe_2022, Harikane_2022, Atek_2023}) and spectroscopic redshift (e.g., \citealt{Schaerer_2022, Roberts_Borsani_2022}). 
Additionally, about $25$ galactic systems hosting a central SMBH have been spectroscopically detected by JWST in the redshift range $4 \lesssim z \lesssim 10$, typically using the broad emission line scaling relations \citep[e.g.,][]{FurtakLabbe2023, GouldingGreene2023, Harikane_2023, Kocevski_2023, Maiolino_2023_new, Ubler_2023}. Recently, \cite{Pacucci_2023_mmstar} and \citet{Maiolino_2023_new} argued that these systems violate the local $\Mblack-M_\star$ relation \citep{Kormendy_Ho_2013, Reines_Volonteri_2015} at $>3\sigma$.
Moreover, \cite{Bogdan_2023} and \cite{Natarajan_2023} reported the X-ray detection of a SMBH at $z=10.3$, with a mass comparable to the stellar mass of its host, and \cite{Pacucci_2022_HD} found that the photometric properties of two galaxy candidates at $z \sim 13$ could be explained by an accreting SMBH of $\sim 10^8 \Msun$.

By pushing the frontier of the farthest quasar detected, we automatically gain insights into the properties of the first population of BHs, or BH seeds \citep{KroupaSubr2020,Pacucci_2022_search, LiInayoshi2023}. As BH seeds should be formed in the redshift range $z \sim 20-30$ \citep{BL01}, detections of higher redshift quasars shorten the time between formation and observation, thus significantly shrinking the uncertainty associated with the size of the parameter space that describes the properties of the first BHs.

Typically, BH seeds are divided into light ($\lesssim 10^3 \Msun$) and heavy ($\gtrsim 10^3 \Msun$) seeds. Light seeds can be formed as remnants of Population III stars or via dynamical processes (e.g., \citealt{PZ_2002, Freitag_2006, Miller_2012, Katz_2015, Lupi_2016, Boekholt_2018,GonzalezKremer2021,ToyouchiInayoshi2023}); conversely, heavy seeds are typically formed from the collapse of pristine gas clouds at very high redshifts (e.g., \citealt{Loeb_Rasio_1994, Bromm_Loeb_2003, Lodato_Natarajan_2006}).

In this study, we expand on the work by \cite{Pacucci_2022_search} by using Bayesian analysis, as well as a far more comprehensive catalog of quasars at $z>6$, to derive the parameters of the distribution of BH seeds. For the first time, we place statistical constraints on the population of BH seeds that accreted gas to form the quasars we observe at redshift $z>5$, which carries crucial insights on the formation of the first BHs in the Universe.

This paper is organized as follows. In Section~\ref{sec:methods}, we discuss our statistical framework to constrain the seed population of BHs. In Section~\ref{sec:results}, we present our results. Finally, in Section~\ref{sec:conclusions}, we discuss the implications of our results and draw our conclusions.

\section{Methods} \label{sec:methods}

In this Section, we outline our method to derive statistical constraints on the population of BH seeds.

\begin{figure} 
\centering
\includegraphics[scale=0.575]{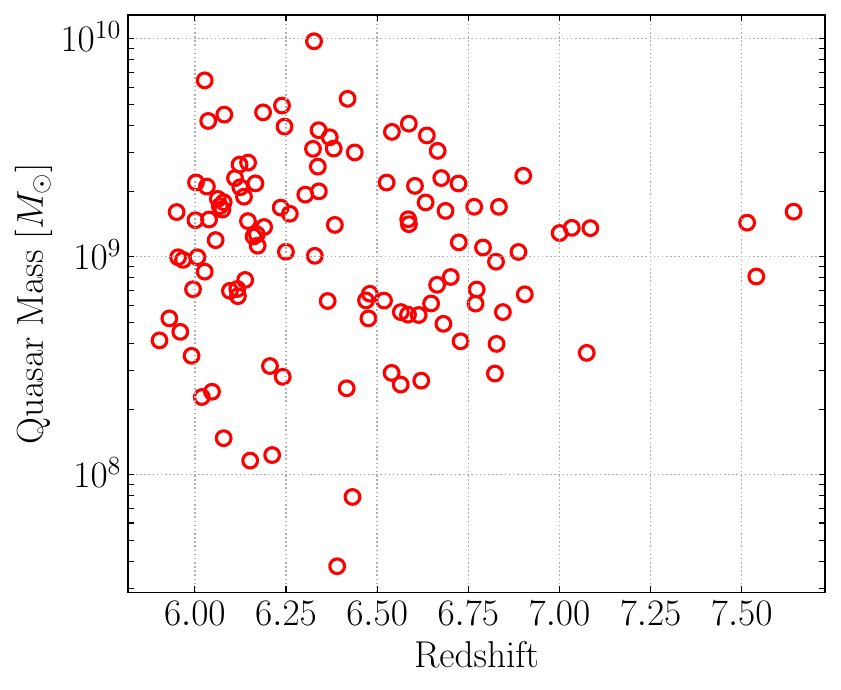}
\caption{Estimated redshifts and masses of the farthest known quasars ($\Mblack \gtrsim 10^8 \Msun$ and $z > 5.9$) from the review by \cite{Fan_2022_review}.}
\label{fig:data}
\end{figure}

We use the recently published \citep{Fan_2022_review} catalog of 113 $z > 5.9$ quasars with robust BH mass estimates from the Mg~II line\footnote{The mass estimate is derived from the widely-used relation presented in \cite{Vestergaard_2009}, based on the Full Width at Half Maximum (FWHM) of the Mg~II line, and the luminosity at $3000 \rm \AA$. As the intrinsic scatter ($\sim 0.55$ dex) dominates the uncertainty for single sources \citep{Fan_2022_review}, this method is robust.}. Their mass and redshift distributions are displayed in Figure~\ref{fig:data}. The Supplementary Material section of \cite{Fan_2022_review} contains a table with detailed information on all the quasars included. The highest-redshift sample (i.e., $z>7$) is derived from \cite{Mortlock_2011, Wang_2018, Banados_2018, Matsuoka_2019, Yang_2019, Yang_2020, Wang_2021}, which are selected using a variety of ground-based and space-based facilities. Note that we have excluded the sample of SMBHs detected by JWST. Specifically, we aimed to work with a consistent sample characterized by common observational selection biases, allowing for uniform modeling across all quasars.

In the catalog, no quasars are detected with an absolute magnitude at $\lambda = 1450\, \rm \AA$ fainter than $\mathcal{M}^{\rm thr}=-24.4$, which we adopt as our detection threshold to model observational completeness. We relate the average absolute magnitude at $\lambda = 1450\, \rm \AA$ to the black hole mass $\Mblack$ and the Eddington ratio $f_{\rm edd}$ (defined as the ratio between the actual mean accretion rate and the mean Eddington accretion rate) via a bolometric correction $C_{1450} = 4.2$ \citep{Runnoe_2012}. This is defined as $L_{\rm bol} = C_{1450} \lambda L_\lambda$. Hence, in the AB magnitude system, we derive the relation
\begin{equation}
    \mathcal{M}^{\rm th} = -3.78 -2.5 \log \left(f_{\rm edd} \frac{\Mblack}{\Msun}\right)\,.
\end{equation}
We note that the bolometric correction indicates the \textit{average} ratio between the bolometric luminosity and the in-band luminosity of a specific class of objects. Of course, the effective ratio for members of that class varies.

The SMBH mass at a cosmic time $t(z)$ (where $z$ is the detection redshift) is related to the initial BH seed mass, $m_{\rm seed}$, by
\begin{equation}
    \Mblack=m_{\rm seed}\exp\left[f_{\rm edd} \mathcal{D} \left(\frac{1-\epsilon}{\epsilon}\right) \left(\frac{t(z) - t_{\rm seed}}{t_{\rm edd}}\right) \right]\,,
    \label{eqn:mbhz}
\end{equation}
where $t_{\rm seed}=130$~Myr is the assumed formation time ($z=25$, see, e.g., \citealt{BL01}), $t_{\rm edd}=450$~Myr is the (Salpeter) growth time scale, $\mathcal{D}$ is the duty cycle (the fraction of time that the BH has accreted since its formation time) and $\epsilon$ is the mean radiative efficiency factor over that time. Note that $f_{\rm edd}$ and $\mathcal{D}$ are degenerate.

Given our set of data of observed absolute magnitudes, $\pmb{\rm d}=\pmb{\mathcal{M}}^{\rm obs}$, the posterior probability on the parameters that describe our model $\pmb{\lambda}$ can be written as
\begin{equation}
    p(\pmb{\lambda}|\pmb{\rm d}) \propto \pi (\pmb{\lambda})\prod_{i} \frac{\int \mathcal{L}({\rm d}_i|\pmb{\theta}) p_{\rm pop}(\pmb{\theta}|\pmb{\lambda}) d\pmb{\theta}}{\int P_{\rm det}(\pmb{\theta}) p_{\rm pop}(\pmb{\theta}|\pmb{\lambda}) d\pmb{\theta}}\,.
    \label{eqn:like}
\end{equation}
In the previous equation, $\pi (\pmb{\lambda})$ is the prior on the parameters $\pmb{\lambda}$ describing the population model $p_{\rm pop}(\pmb{\theta}|\pmb{\lambda})$, $\mathcal{L}({\rm d}_i|\pmb{\theta})$ is the likelihood of observing the data set given the population properties, and $P_{\rm det}(\pmb{\theta})$ is the fraction of events in the Universe that would be detected for a particular population model, characterized by the population parameters $\pmb{\lambda}$. In our analysis, we have $\pmb{\lambda}=\{f_{\rm edd}, \pmb{\Upsilon}, \pmb{\Xi}\}$, where $\pmb{\Upsilon}=\{\mathcal{D}, \epsilon \}$ and $\pmb{\Xi}$ is the set of parameters describing the shape of the distribution of $m_{\rm seed}$.

\begin{figure*} 
\centering
\includegraphics[scale=0.45]{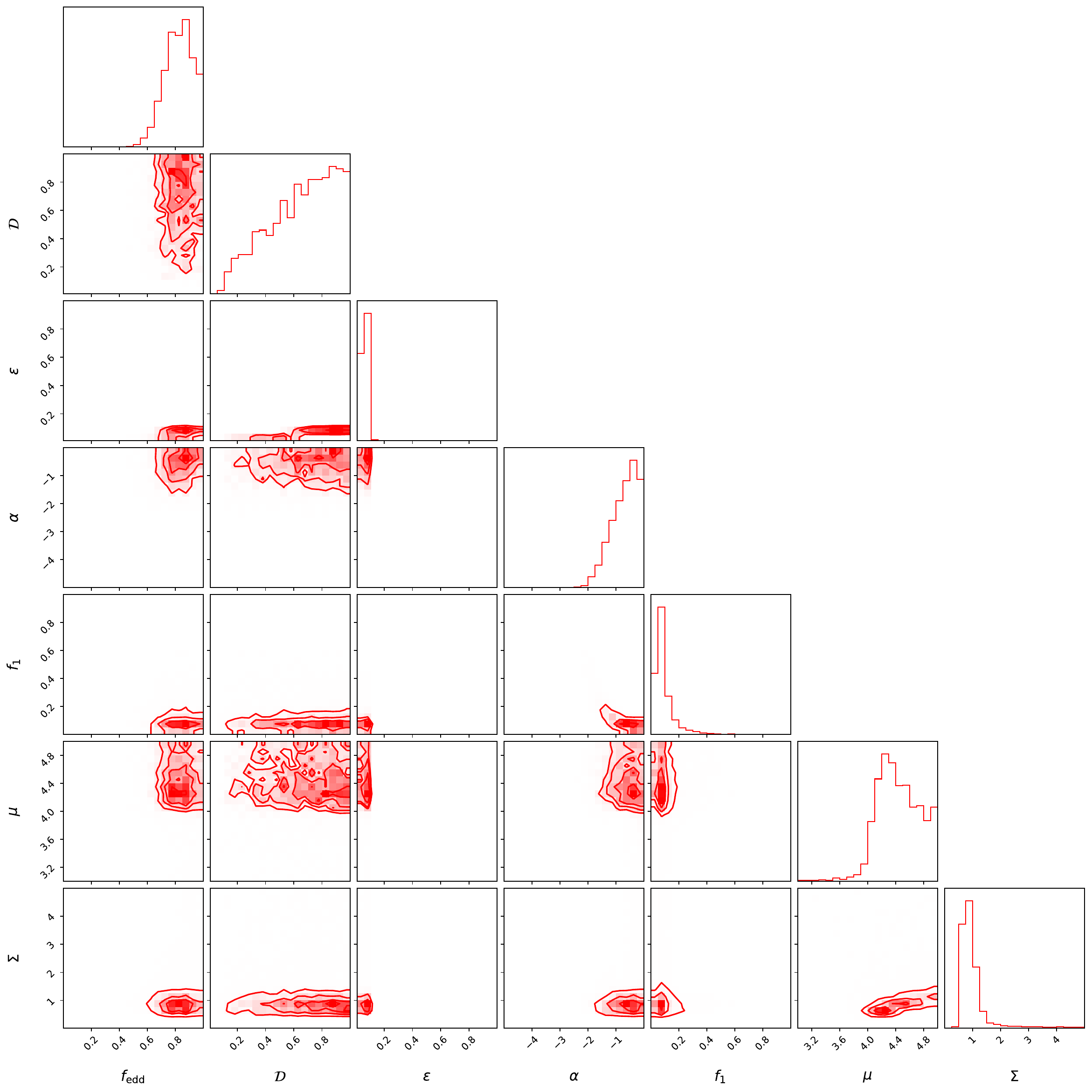}
\caption{Hyper-posterior for the parameters describing the accretion of SMBHs (Eq.~\ref{eqn:mbhz}) and the mass distribution of BH seeds (Eq.~\ref{eqn:seeds}) for our model ``Power Law + Lognormal'', representative of forming heavy seeds in the mass range $[10^3-10^5]\Msun$ following a lognormal distribution, and light seeds in the mass range $[10-10^3]\Msun$ following a power-law distribution.}
\label{fig:corner}
\end{figure*}

We model the likelihood of observing the data set given the population properties as
\begin{equation}
    \mathcal{L}({\rm d}_i|\pmb{\theta}) = \frac{1}{\sigma_{i}\sqrt{2\pi}}\exp\left[-\frac{(\mathcal{M}_{\rm i}^{\rm obs}-\mathcal{M}^{\rm th})^2}{2\sigma_{i}^2}\right]\,,
\end{equation}
where $\sigma_i$ represents the uncertainty in the measurement, whose upper limit we conservatively take to be $10\%$, to agree with the uncertainty reported in \cite{Fan_2022_review}. This error accounts for uncertainties in the determination of the absolute magnitude of the source, determined from the relative magnitude, which is estimated with great accuracy, and the redshift; in the case of Mg~II line redshift measurements, typical estimates are well within the $0.5\%$ of the true value. 

By definition, the fraction of events in the Universe that would be detected for a particular population model, characterized by the population parameters, is
\begin{eqnarray}
    P_{\rm det}(\pmb{\theta}) &=& \int_{\mathcal{M}^{\rm obs}<\mathcal{M}^{\rm thr}} \mathcal{L}({\rm d}_i|\pmb{\theta}) d {\rm d}_i \nonumber\\
    &=& \frac{1}{2}\left[1+{\rm erf}\left(\frac{\mathcal{M}_{\rm i}^{\rm obs}-\mathcal{M}^{\rm th}}{\sigma_i\sqrt{2}}\right)\right]\,.
\end{eqnarray}
This quantity represents the number of events that would pass a threshold and, therefore, the completeness of the observed sample. In our case, any SMBH with an absolute magnitude larger than $\mathcal{M}^{\rm thr}$ is undetected. Finally, the population model is taken to be
\begin{eqnarray}
    p_{\rm pop}(\pmb{\theta}|\pmb{\lambda}) &=& p(\mathcal{M}^{\rm th}|\Mblack,f_{\rm edd})\nonumber\\
    &\times& p(\Mblack|f_{\rm edd},\pmb{\Upsilon},m_{\rm seed})p(m_{\rm seed}|\pmb{\Xi})\,,
\end{eqnarray}
whose characteristics are described by the set of parameters $\pmb{\Xi}$.

Since BH seeds come in two flavors (see Section \ref{sec:intro}), we take the mass distribution to be described by the sum of two components (``Power Law + Lognormal'' - PLN), which are reminiscent of the theoretical distributions of light seeds (``Power Law'' - PL) and heavy seeds (``Lognormal'' - LN)
\begin{eqnarray}
    p(m_{\rm seed}|\pmb{\Xi}) &=& f_1 m_{\rm seed,PL}(\pmb{\Xi}_{\rm PL}) \nonumber\\
    & \times& H(m_{\rm seed}-m_{\rm seed,PL}^{\rm min})H(m_{\rm seed,PL}^{\rm max}-m_{\rm seed})\nonumber\\
    &+& (1-f_1)m_{\rm seed, LN}(\pmb{\Xi}_{\rm LN})H(m_{\rm seed}-m_{\rm seed,LN}^{\rm min})\nonumber \\
    &\times& H(m_{\rm seed,LN}^{\rm max}-m_{\rm seed})\,.
    \label{eqn:seeds}
\end{eqnarray}
In the previous equation, $\pmb{\Xi}=\left\{\pmb{\Xi}_{PL}, \pmb{\Xi}_{LN}, f_1 \right\}$, $H$ is the Heaviside function, $\{m_{\rm seed,PL}^{\rm min},m_{\rm seed,PL}^{\rm max}\}$ are the minimum and maximum seed masses for the PL distribution, $\{m_{\rm seed,LN}^{\rm min},m_{\rm seed,LN}^{\rm max}\}$ are the minimum and maximum seed masses for the LN distribution, which can have values in the range $[10-10^5]\Msun$, and
\begin{equation}
    m_{\rm seed,PL}(\pmb{\Xi}_{\rm PL}) = m_{\rm seed}^{-\alpha}
\end{equation}
\begin{equation}
    m_{\rm seed,LN}(\pmb{\Xi}_{\rm LN}) = \frac{1}{m_{\rm seed} \Sigma \sqrt{2\pi}}\exp\left[-\frac{(\log m_{\rm seed}-\mu)^2}{2\Sigma^2}\right]\,,
\end{equation}
with $\pmb{\Xi}_{PL}=\left\{\alpha\right\}$ and $\pmb{\Xi}_{LN}=\left\{\mu, \Sigma \right\}$.

In our analysis, we use the following priors: uniform in the range $[0,1]$ for $f_{\rm edd}$ (i.e., we do not allow mean accretion rates that are super-Eddington), uniform in the range $[0-1]$ for  $\mathcal{D}$, and uniform in the range $[0.01-1]$ for $\epsilon$. For the parameters that described the BH mass function, we have priors that are uniform and in the range $[0,1]$ for $f_1$, $[-5,0]$ for $\alpha$, $[3,5]$ for $\mu$, and $[0,3]$ for $\Sigma$.

\begin{figure} 
\centering
\includegraphics[scale=0.575]{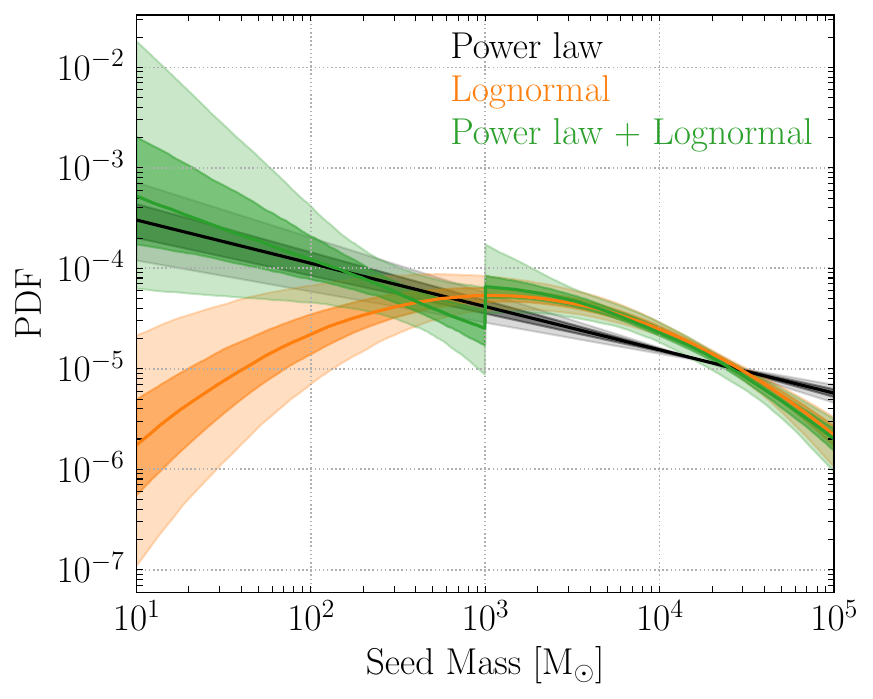}
\caption{Inferred mass distribution of BH seeds in the case of a ``Power Law + Lognormal'' (green), ``Power Law'' (black), and ``Lognormal'' (orange) function. The solid line corresponds to the median at each mass, while shaded bands denote $50$ per cent and $90$ per cent credible intervals.}
\label{fig:mdist}
\end{figure}

\begin{table*}
\centering
\begin{tabular}{lcccccccc}
\hline\hline
Model & $f_{\rm edd}$ & $\mathcal{D}$ & $\epsilon$ & $\alpha$ & $\mu$ & $\Sigma$ & $f_1$ & $\log Z$\\
\hline
Power Law + Lognormal & $0.82^{+0.10}_{-0.10}$ & $0.66^{+0.23}_{-0.23}$ & $0.06^{+0.02}_{-0.02}$ & $-0.70^{+0.46}_{-0.46}$ & $4.44.^{+0.30}_{-0.30}$ & $1.02^{+0.60}_{-0.60}$ & $0.09^{+0.07}_{-0.07}$ & $-200.4$ \\
\hline\hline
Power Law & $0.82^{+0.10}_{-0.10}$ & $0.67^{+0.23}_{-0.23}$ & $0.06^{+0.02}_{-0.02}$ & $-0.43^{+0.07}_{-0.07}$ & -- & -- & -- & $-206.2$ \\
Lognormal & $0.83^{+0.10}_{-0.10}$ & $0.68^{+0.23}_{-0.23}$ & $0.07^{+0.02}_{-0.02}$ & -- & $4.44^{+0.24}_{-0.24}$ & $0.79^{+0.11}_{-0.11}$ & -- & $-197.4$ \\
\hline\hline
Power Law ($[10-10^3]\Msun$) & $0.99^{+0.01}_{-0.01}$ & $0.67^{+0.22}_{-0.22}$ & $0.06^{+0.02}_{-0.02}$ & $-0.42^{+0.09}_{-0.09}$ & -- & -- & -- & $-971.1$ \\
Lognormal ($[10^3-10^5]\Msun$) & $0.98^{+0.02}_{-0.02}$ & $0.65^{+0.23}_{-0.23}$ & $0.08^{+0.03}_{-0.03}$ & -- & $4.78^{+0.14}_{-0.14}$ & $0.72^{+0.08}_{-0.08}$ & -- & $-355.6$ \\
\hline\hline
\end{tabular}
\caption{Summary of the results for various initial mass functions of BH seeds. The first column lists the models' names, followed by their parameters' inferred values. The last column reports the log evidence of the model.}
\label{tab:models}
\end{table*}

\section{Results} \label{sec:results}

Using the data on the farthest quasars from \citet{Fan_2022_review}, we fit their masses as a function of the parameters that describe their accretion ($f_{\rm edd}, \mathcal{D}, \epsilon$) and of the set of parameters $\pmb{\Xi}$ that describe the mass distribution of BH seeds. We use the nested-sampling code \textsc{nestle}\footnote{\url{http://kylebarbary.com/nestle/index.html}} to maximize the log-likelihood of our model and to infer the confidence regions of our parameters. Notably, the nested algorithm also supplies us with the marginalized likelihood, which can be used for model selection.

For our primary model, we assume that the mass distribution of BH seeds is described by the sum of power-law and lognormal distributions, with $\{m_{\rm seed,PL}^{\rm min},m_{\rm seed,PL}^{\rm max}\}=\{10,10^3\}\Msun$ and $\{m_{\rm seed,LN}^{\rm min},m_{\rm seed,LN}^{\rm max}\}=\{10^3,10^5\}\Msun$. This distribution is likely the most accurate from a physical point of view, aligning with the typical predicted form of the distributions for light and heavy BH seeds within their respective mass categories (see, e.g., \citealt{Volonteri_2010, Ferrara_2014}, and the discussion in \citealt{Pacucci_2022_search}).

We show the hyper-posterior distribution of the parameters that describe the mass distributions of the BH seeds, along with the inferred values of the parameters that describe their growth, in Figure~\ref{fig:corner}. The inferred values of the Eddington fraction, the duty cycle, and the mean radiative efficiency are $0.82^{+0.10}_{-0.10}$, $0.66^{+0.23}_{-0.23}$, and $0.06^{+0.02}_{-0.02}$, respectively. We find that the slope of the power-law portion of the mass distribution is $-0.70^{+0.46}_{-0.46}$, while the mean and variance of the lognormal portion are $4.44.^{+0.30}_{-0.30}$ and $1.02^{+0.60}_{-0.60}$.

For comparison, we consider the case that the mass distribution of BH seeds is described by a simple power law (in this case, we fix $f_1=1$) over the whole mass range $[10-10^5]\Msun$. We also use another form of the distribution, a pure lognormal (in this case, we fix $f_1=1$) over the whole mass range $[10-10^5]\Msun$. Table~\ref{tab:models} reports the parameters that describe the mass distributions of the BH seeds and the inferred values of the parameters that describe their growth. We observe that the Eddington fraction, the duty cycle value, and the mean radiative efficiency remain consistent throughout our models.

Figure~\ref{fig:mdist} shows the inferred mass distribution of BH seeds in the three cases described above. 
We observe that the PLN distribution exhibits a clear overlap with the LN distribution for masses greater than $\sim 10^3\Msun$, while it transitions towards a power-law distribution for masses below $\sim 10^3\Msun$, albeit with a gentler slope.

We also report the results for two more models, where we restrict the PL model to BH seed masses in the range $[10,10^3]\Msun$ and the LN model to BH seed masses $[10^3,10^5]\Msun$, reminiscent of the typical mass range of light and heavy seeds discussed in the literature. These models prefer Eddington fractions close to unity but with the duty cycle value and the mean radiative efficiency consistent with the other models. 

In Table~\ref{tab:models}, we also report the log evidence ($\log Z$) of the three models, defined as the marginalized likelihood in Eq.~\ref{eqn:like}. The log evidence serves as a tool for conducting a Bayesian ratio test to determine the preferable model. The model with the highest log evidence is favored in this test. Given the values of their log evidence, the PL model restricted to the range $[10,10^3]\Msun$ and the LN model restricted to the range $[10^3,10^5]\Msun$ are very disfavored statistically with respect to our reference model that represents the typical predicted form of the distributions for light and heavy BH seeds within their respective mass categories.

\section{Discussion and Conclusions}
\label{sec:conclusions}

In this Letter, we developed a state-of-the-art Bayesian analysis to infer the mass distribution properties of BH seeds that originated the quasars we observe in the high-redshift Universe. By combining data on their redshift and mass, derived from the review by \cite{Fan_2022_review}, we have obtained accurate constraints on the properties of the seed population.

We have shown that the distribution of BH seeds' masses can be best characterized by the combination of a power law and a lognormal function within the mass intervals of $[10-10^3]\Msun$ and $[10^3-10^5]\Msun$, respectively. This combination corresponds appropriately to the light and heavy seeds. Our analysis yielded a power-law slope of $-0.70^{+0.46}_{-0.46}$ and a lognormal mean of $4.44^{+0.30}_{-0.30}$. Models that exclusively incorporate either a power law or a lognormal function within the respective mass ranges for light and heavy seeds are statistically strongly disfavored. This implies that both components are necessary to explain the mass distribution of high-redshift quasars.

Before discussing the implications of our findings, we note that Eq.~\ref{eqn:mbhz} describes the \textit{average} growth process of black hole seeds. For instance, $f_{\rm edd}$ therein is an average value (from formation time to observation) and we do not model accretion histories with consistent super-Eddington accretion, which may affect our results. For example, \cite{Maiolino_2023} estimates a significantly super-Eddington rate for GN-z11. However, in some cases, the instantaneous values of $f_{\rm edd}$ detected by observations may be significantly different than the average value. Finally, the value of $t_{\rm seed}$ could have a different value for different seeding processes, while we kept it fixed in our analysis to reduce the computational cost of our statistical analysis. While we do not make any specific assumptions on the channels that produce our BH seeds, we note that our chosen value of $t_{seed}=130$\,Myr (i.e., $z=25$) is appropriate for BH seeds that are formed either as remnants of Pop III stars \citep{BL_2000} or as direct collapse, heavy BH seeds \citep{Yue_2014_brief_era}. We defer the examination of the impact of these changes on future research.

Constraining the properties of the population of BH seeds, such as $f_{\rm edd}$, $\mathcal{D}$, $\epsilon$, and the parameters describing the shape of the distribution of $m_{\rm seed}$, is crucial for a variety of astrophysical and cosmological applications \citep[e.g.,][]{Izquierdo-VillalbaBonoli2020,SpinosoBonoli2023}. For example, most large-scale cosmological simulations, e.g., ASTRID \citep{Ni_2022, Bird_2022} and IllustrisTNG \citep{Springel_2018, IllustrisTNG_2018, Nelson_2019_Illustris} include active galactic nuclei feedback, with significantly different seeding prescriptions. Recent studies have shown that seeding prescriptions profoundly impact the evolution of individual galaxies (e.g., \citealt{Weinberger_2017, Impact_cosmo_sim_2019}).

The mass distributions of BH seeds are in the mass range of the elusive population of intermediate-mass BHs. The attributes of these distributions, along with their distribution across different redshifts, have a crucial role in influencing the rates and characteristics of gravitational wave detections using upcoming observatories. Both the Laser Interferometer Space Antenna (LISA, \citealt{LISA_2023}) for BH masses $\gtrsim 10^3 \Msun$ and the third-generation, ground-based gravitational wave detectors, with an emphasis on low-frequency and thus for masses $\lesssim 10^3 \Msun$ (Cosmic Explorer, e.g., \citealt{ReitzeAdhikari2019} and Einstein Telescope, e.g., \citealt{PunturoAbernathy2010}) are predicted to detect this population of high-$z$ sources systematically, and further constrain their properties \citep{Pacucci_2020,Chen_Hsin_2022,FragioneLoeb2023}.

For the first time, our study has introduced a framework that enables us to leverage the currently detected quasars, while properly modeling their observational completeness, to infer the properties of BH seeds in the early Universe. As new, more distant quasars become known, especially with JWST \citep[see, e.g.,][]{Larson_2023, Maiolino_2023}, our constraints will become more robust and more descriptive of this, thus far, elusive population of ancient BHs.

\section*{Acknowledgements}
G.F. acknowledges support by NASA Grant 80NSSC21K1722 and NSF Grant AST-2108624 at Northwestern University. F.P. acknowledges support from a Clay Fellowship administered by the Smithsonian Astrophysical Observatory. This work was also supported by the Black Hole Initiative at Harvard University, which is funded by grants from the John Templeton Foundation and the Gordon and Betty Moore Foundation.

\bibliography{ms}{}
\bibliographystyle{aasjournal}

\end{document}